\documentclass[preprint,onecolumn,showpacs,preprintnumbers,amsmath,amssymb]{revtex4}

\usepackage{mathrsfs}
\usepackage{epsfig}
\usepackage{amsmath}
\usepackage{subfigure}

\newcommand{\bs}{\boldsymbol}
\newcommand{\be}{\begin{equation}}
\newcommand{\ee}{\end{equation}}
\newcommand{\bea}{\begin{eqnarray}}
\newcommand{\eea}{\end{eqnarray}}
\newcommand{\nn}{\nonumber}

\begin{document}

\title{High frequency susceptibility of a weak ferromagnet with magnetostrictive magnetoelectric coupling: using heterostructures to tailor electromagnon frequencies}

\author{K.~L.~Livesey}
\email{livesey@physics.uwa.edu.au}
\author{R.~L.~Stamps}

\affiliation{School of Physics M013, University of Western Australia, 35 Stirling Hwy, Crawley WA 6009, Australia}

\date{\today}

\begin{abstract}
In the first part of this work we calculate the high frequency magnetoelectric susceptibility of a simultaneously ferroelectric and canted antiferromagnetic (also know as weak ferromagnetic) thin film with magnetostrictive magnetoelectric coupling. We show that a dynamic coupling exists between the ferroelectric and optic antiferromagnetic excitations. In the second part of the paper, we calculate using an effective medium method the susceptibility of a heterostructure comprising alternating thin films of such a material together with a ferromagnet. Dipolar magnetic fields serve to couple the ferromagnetic and optic antiferromagnetic modes, which in turn couples the ferromagnetic and ferroelectric excitations. This provides a mechanism for creating ``electromagnon" modes in the microwave regime which may be useful for applications.
\end{abstract}


\maketitle

\section{Introduction}
\label{Introduction}

In the 1970s, Bar'yakhtar and Chupis calculated the high frequency magnetic, electric and magnetoelectric susceptibility of a model ferroelectric ferromagnet using second quantization of the electric polarization $\bs{P}$ and the magnetization $\bs{M}$ fields \cite{BarChupis}. They noted that the equilibrium directions of $\bs{P}$ and $\bs{M}$ must not be parallel or perpendicular in order for there to be a dynamic magnetoelectric coupling and in order for the existence of coupled excitations, known as ``electromagnons." Maugin found a similar result \cite{Maugin81}. However, most simultaneously magnetic and ferroelectric materials (known as multiferroic) are not ferromagnetic and have a more complicated spin structure.

Later, in 1982, Tilley and Scott used a Landau-Ginzburg free energy and equations of motion to calculate the full high frequency susceptibility of the antiferromagnetic dielectric BaMnF$_4$ \cite{TilleyScott}. They were able to explain the observed frequency dependent dielectric anomaly in BaMnF$_4$ by including a magnetoelectric coupling term of the form $(\beta_{1} p + \beta_{2} p^2) M^{x} L^{z}$, where $p$ is the dielectric polarization, $\bs{M}=\bs{M}_{a}+\bs{M}_{b}$ and $\bs{L}=\bs{M}_a - \bs{M}_{b}$. The subscripts $a$ and $b$ denote the two antiferromagnetic sublattices. $\beta_{1}$ and $\beta_{2}$ give the strengths of the magnetoelectric coupling. This magnetoelectric coupling term is a Dzyaloshinskii-Moriya type term \cite{DzyaloWFM,Moriya} that causes a canting of the antiferromagnetic sublattices (hence the material is called a weak ferromagnet) that may be altered by application of an electric field \cite{DzyaloME}.

A similar term has been used to model weak ferromagnetic BiFeO$_3$, the only known room temperature magnetoelectric multiferroic material. deSousa and Moore demonstrated how a coupling, $\bs{P} \cdot \bs{M}_{a} \times \bs{M}_{b}$, could lead to electric field control of magnon dispersion with potential applications to spin wave logic devices \cite{desousaSW}.

This type of coupling in BiFeO$_3$ now seems unlikely since there is no observation of a change in the weak ferromagnetic moment when $\bs{P}$ is reversed or when an electric field is applied \cite{zhao06,Ederer05}. However, a magnetoelectric coupling energy which is always symmetry allowed is \cite{smol}
\be
\mathscr{E} = (J + \Gamma P^2 ) \bs{M}_{a} \cdot \bs{M}_{b} ,
\ee
where $J$ is the antiferromagnetic exchange constant which is perturbed slightly by the influence of the electric polarization $P$. Ionic distortions in the displacive ferroelectric cause changes to the effective exchange interaction between spins, giving rise to this magnetostrictive magnetoelectric coupling. We will show that this type of coupling, together with a weak canting of antiferromagnetic sublattices (so that $\bs{M}_{a/b}$ and $\bs{P}$ are not exactly perpendicular), allows for a dynamic magnetoelectric coupling. It is the ``optic" antiferromagnetic mode, where the two antiferromagnetic sublattices oscillate out-of-phase, which hybridizes with the dielectric mode.

In Sec.~\ref{SinglePhase} we calculate the full frequency-dependent magnetoelectric susceptibility tensor analytically from a starting free energy for a thin film ferroelectric weak ferromagnet. Components of the susceptibility tensor have poles at the magnetic, electric and magnetoelectric resonant frequencies.

In Sec.~\ref{heterostructure} we extend our calculation to consider a heterostructure containing alternating ferromagnetic and ferroelectric weak ferromagnetic layers. Using a particular effective medium method for long wavelength dielectric excitations \cite{agranovich} and long wavelength magnetic excitations \cite{raj87, almeida88}, the susceptibility is found analytically and reduces to known limits. We find that the ferromagnetic resonance couples to the optic antiferromagnetic mode via dipolar fields and hence also couples to the dielectric mode in the weak ferromagnet. Therefore there is a magnetoelectric resonance in the low GHz (microwave) regime, whereas in single-phase magnetoelectric materials the resonances are usually all in the infrared regime. By combining such a ferromagnet and weak ferromagnet it is possible to create an effective material for tuning magnetoelectric response frequencies. Applications exist to microwave signal processing using applied electric fields \cite{fetisov05shifter,ustinov} or even to designing left-handed materials in small frequency ranges \cite{Liu05}.

\section{Weak ferromagnet susceptibility}
\label{SinglePhase} 

\subsection{Geometry and energy density}

\begin{figure}[h]
\begin{center}
\includegraphics[width=8cm]{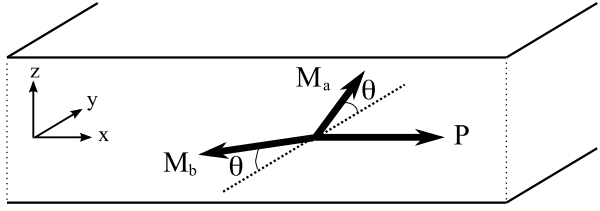}
\caption{\label{WFMgeom}The ferroelectric weak ferromagnet geometry.}
\end{center}
\end{figure}

The geometry of the ferroelectric weak ferromagnet is shown in Fig.~\ref{WFMgeom}. The electric polarization $\bs{P}$ lies along the $x$ direction. The two antiferromagnetic sublattices $\bs{M}_{a/b}$ lie perpendicular to the electric polarization, predominantly along the $y$ direction, but are canted in the $z$ direction by an angle $\theta$. This angle is exaggerated in Fig.~\ref{WFMgeom} and in BiFeO$_3$, for example, is given by $\theta \sim 0.14^{\circ}$ \cite{bea07}. When considering the thin film geometry, the film thickness is in the $z$ direction. This minimizes depolarizing plus demagnetizing energies.
 
 The energy density of the antiferromagnet is given by:
 \bea
\mathscr{E}_{AFM} &=& (J+ \Gamma (P^{x})^{2} ) \bs{M}_{a} \cdot \bs{M}_{b} - K \left[ \left(M_{a}^{y} \right)^2 + \left( M_{b}^{y} \right)^2  \right] 
\nn \\
&&+ D (M_{a}^{y} M_{b}^{z} - M_{a}^{z} M_{b}^{y} ) - \bs{h} \cdot \bs{M} + 2 \pi (M_{a}^{z} + M_{b}^{z})^{2}.
\label{newAFMenergy}
\eea
The first term represents the antiferromagnetic exchange interaction with $J >0$. A weak contribution to the exchange constant is due to so-called ``isotropic" or magnetostrictive magnetoelectric coupling with strength given by $\Gamma$ \cite{smol}. It arises since the soft-phonon mode associated with the electric polarization in the film $\bs{P}$ is coupled to the magnetic system through magnetostriction. The second term is a uniaxial anisotropy energy which favors the alignment of the sublattice magnetization in the $y$ direction. The third term in Eq.~\eqref{newAFMenergy} is the Dzyaloshinskii-Moriya interaction with a Dzyaloshinskii-Moriya vector $\bs{D} = D \hat{\bs{x}}$ giving canting in the $z$~direction. The fourth term describes the interaction with a small driving field $\bs{h}$. Finally, the last term in Eq.~\eqref{newAFMenergy} is the demagnetizing term in CGS form, assuming that the thin film geometry has an interface containing both sublattices.

The energy density of the dielectric part of the system is given by:
\bea
\mathscr{E}_{FE} &=& - \frac{1}{2} \xi \left( P^{x} \right)^2 + \frac{1}{4} \Delta \left( P^{x} \right)^4  - \bs{e} \cdot \bs{P}  \nn \\
&&+ \frac{1}{2} \xi_{\perp} \left[ \left( P^{y} \right)^2 + \left( P^{z} \right)^2 \right] +2 \pi \left( P^{z} \right)^2 .
\label{E_FElayer}
\eea
$\xi$ and $\Delta$ are phenomenological Landau coefficients giving a spontaneous polarization in the $x$~direction. The isotropic magnetoelectric coupling constant $\Gamma$ (see Eq.~\eqref{newAFMenergy}) alters $\xi$ by a small amount. A one-dimensional model for the spontaneous polarization is valid when examining small amplitude dynamics about equilibrium. The third term in Eq.~\eqref{E_FElayer} is the interaction of the dielectric with a small driving field $\bs{e}$. The fourth term describes the strength of the dielectric stiffness $\xi_{\perp}>0$ of the material in the $y$ and $z$ directions. We make a simplifying assumption that the system is isotropic in the $y$-$z$ plane. The last term is the depolarizing energy density.

  \subsection{Equations of motion}
 
From the free energy, the equations of motion for the magnetization and polarization can be found using the Landau-Lifshitz (or torque) equation and the Landau-Khalatnikov relaxation equation respectively
\bea
\frac{ d \bs{M} }{ d t} &=& \gamma \bs{M} \times \left( - \frac{\partial \mathscr{E}}{\partial \bs{M}}  \right)  ,
\label{LLequation} \\
\frac{ d^2 \bs{P} }{ d t^2  } &=& f \left( - \frac{\partial \mathscr{E} }{\partial \bs{P} } \right) ,
\label{Peqnmotion}
\eea
where the derivatives, $ - \frac{\partial \mathscr{E}}{\partial \bs{M}} $ and $- \frac{\partial \mathscr{E} }{\partial \bs{P} }$, represent the effective magnetic field and the effective electric field acting on the systems. $\gamma$ is the gyromagnetic ratio and $f$ is an effective inverse mass term for the dielectric oscillations. We ignore damping in both equations.

The equations of motion are obtained by substituting Eqs.~\eqref{newAFMenergy} and \eqref{E_FElayer} into Eqs.~\eqref{LLequation} and \eqref{Peqnmotion}, assuming oscillating solutions that vary in time according to $e^{-i \omega t}$, and then linearizing the resulting equations. The linearization is done by splitting the two sublattice magnetizations and the polarization into static and small dynamic parts and then ignoring terms which are quadratic in small dynamic terms. If dynamic parts are denoted by lower case letters, then according to the geometry shown in Fig.~\ref{WFMgeom} the equations are linearized using:
\bea
\bs{M}_{a} &=& (a_{}^{x}, M_{0} \cos \theta + a_{}^{y}, M_{0} \sin \theta + a_{}^{z} ),
\label{MaLinearise} \\
\bs{M}_{b} &=& (b_{}^{x}, -M_{0} \cos \theta + b_{}^{y}, M_{0} \sin \theta + b_{}^{z} ),
\label{MbLinearise} \\
\bs{P} &=& (P_{0} + p^{x}, p^{y}, p^{z} ) .
\label{P2linear}
\eea
The equilibrium canting angle $\theta$ is given by minimizing Eq.~\eqref{newAFMenergy}:
\be
\theta = \frac{1}{2} \arctan \left(  \frac{ D }{ J + \Gamma P_{0}^{2} +K+ 4 \pi }  \right),
\label{ThetaEquilib}
\ee 
and the equilibrium polarization in the $x$~direction, $P_{0}$, is given by minimizing Eq.~\eqref{E_FElayer} plus Eq.~\eqref{newAFMenergy}: 
\be
P_{0} = \sqrt{ \frac{ \left( \xi-2 \Gamma M_{0}^{2} [-\cos^2 \theta + \sin^2 \theta] \right) }{ \Delta }  }.
\label{Pequilib}
\ee

Combining Eqs.~\eqref{newAFMenergy}-\eqref{P2linear} we obtain magnetization equations:
\bea
	-\frac{i \omega}{\gamma} a_{}^{x} &=& -a_{}^{y} \big( [2 H_{d} + H_{ex}(P_{0})+H_{a}] \sin \theta  + H_{DM} \cos \theta \big)  
\nn \\
&&- a_{}^{z} \big( [H_{d}+H_{ex}(P_{0}) + H_{a} ] \cos \theta - H_{DM} \sin \theta \big) 
\nn \\
&&+ b_{}^{y} \big( H_{DM} \cos \theta + H_{ex}(P_{0}^{}) \sin \theta \big)  -b_{}^{z} \big( [H_{d}+H_{ex} (P_{0}^{})  ] \cos \theta - H_{DM} \sin \theta \big)  
\nn \\
&& -p_{}^{x} 2 \Gamma P_{0} M_{0}^{2} \sin (2 \theta)+ M_{0}^{} h_{}^{z} \cos \theta -M_{0}^{} h_{}^{y} \sin \theta
\label{AxEquation2} \\
	-\frac{i \omega}{\gamma} a_{}^{y} &=&  a_{}^{x} \big(  [2 H_{d} + H_{ex}(P_{0}) ] \sin \theta +H_{DM} \cos \theta    \big) -b_{}^{x} H_{ex}(P_{0}) \sin \theta 
\nn \\
&& +M_{0}  h_{}^{x}  \sin \theta 
\label{AyEquation2} \\
	-\frac{i \omega}{\gamma} a_{}^{z} &=& a_{}^{x} \big( [H_{ex}(P_{0}) + H_{a}] \cos \theta - H_{DM} \sin \theta \big) +b_{}^{x} H_{ex}(P_{0}) \cos \theta \nn \\
&& -  M_{0} h_{}^{x} \cos \theta
\label{AzEquation2} \\
	-\frac{i \omega}{\gamma} b_{}^{x} &=& a_{}^{y} \big( H_{ex}(P_{0}) \sin \theta +H_{DM} \cos \theta \big) + a_{}^{z} \big( \left[H_{d}+H_{ex}(P_{0}) \right] \cos \theta - H_{DM} \sin \theta \big)
\nn \\
&& - b_{}^{y} \big( [2 H_{d} +H_{ex}(P_{0}) +H_{a} ] \sin \theta + H_{DM} \cos \theta  \big)
\nn \\
&& + b_{}^{z} \big( [H_{d}+H_{ex}(P_{0}) + H_{a}] \cos \theta - H_{DM} \sin \theta   \big)
\nn \\
&& + p_{}^{x} 2 \Gamma P_{0} M_{0}^{2} \sin (2 \theta) - M_{0} h_{}^{z} \cos \theta - M_{0} h_{}^{y} \sin \theta
\label{BxEquation2} \\
	-\frac{i \omega}{\gamma} b_{}^{y} &=&  - a_{}^{x} H_{ex}(P_{0}) \sin \theta + b_{}^{x} \big( [ 2 H_{d} + H_{ex}(P_{0}) ] \sin \theta + H_{DM} \cos \theta \big)
\nn \\
&& + M_{0} h_{}^{x} \sin \theta 
\label{ByEquation2} \\
	-\frac{i \omega}{\gamma} b_{}^{z} &=& - a_{}^{x} H_{ex} \cos \theta 
- b_{}^{x} \big( [H_{ex}(P_{0}) + H_{a}] \cos \theta - H_{DM} \sin \theta \big)   \nn \\
&&+ M_{0} h_{}^{x} \cos \theta ,
\label{BzEquation2}
\eea
where the effective exchange, anisotropy, demagnetizing and Dzyaloshinskii-Moriya magnetic fields are given respectively by $H_{ex}(P_{0}) = M_{0} (J + \Gamma P_{0}^{2} )$, $H_{a} = 2 K M_{0}$, $H_{d}=4 \pi M_{0}$ and $H_{DM} = M_{0} D$. We write $H_{ex}(P_{0})$ as $H_{ex}$ below to shorten the notation.

The only component of the dielectric polarization to couple with the magnetization equations is $p_{}^{x}$. It's equation of motion is given by:
\bea
- \frac{\omega^2}{f} p_{}^{x} &=& \left( \xi-2 \Gamma M_{0}^{2} [- \cos^2 \theta + \sin^2 \theta] - 3 \Delta P_{0}^2 \right) p_{}^{x} + e_{}^{x} 
\nn \\
&& - 2 \Gamma P_{0} M_{0} \left(  - a_{}^{y} \cos \theta  +  b_{}^{y} \cos \theta + a_{}^{z} \sin \theta + b_{}^{z} \sin \theta   \right) .
\label{P2xEquation}
\eea
It can be seen that if the canting were to vanish, then $\theta, a_{}^{y}, b_{}^{y} \to 0$, and the magnetic and dielectric equations of motion would not be coupled. So although the magnetoelectric coupling enters into the exchange interaction, rather than the Dzyaloshinskii-Moriya interaction, it results in a dynamic magnetoelectric coupling.

The equation of motion for the remaining two components of the dielectric polarization are:
\bea
- \frac{\omega^2}{f} p_{}^{y} &=& - \xi_{\perp} p_{}^{y} + e_{}^{y} 
\label{Pyequation} \\
- \frac{\omega^2}{f} p_{}^{z} &=& - ( \xi_{\perp}+4 \pi )  p_{}^{z} + e_{}^{z} .
\label{Pzequation}
\eea

  \subsection{Susceptibility}
  
  The seven equations of motion Eqs.~\eqref{AxEquation2}-\eqref{P2xEquation} can be used to solve for $\left\{ a_{}^{x}, a_{}^{y}, a_{}^{z}, b_{}^{x}, b_{}^{y}, b_{}^{z}, p_{}^{x} \right\}$ analytically as a function of driving fields $\bs{h}$ and $e_{}^{x}$. First we set $h_{}^{z} \ne 0$ and $h_{}^{x}=h_{}^{y}=e_{}^{x}=0$. Then the equations for $\bs{a}$ and $\bs{b}$ are symmetric under the transformation $b_{}^{x} \to -a_{}^{x}$, $b_{}^{y} \to - a_{}^{y}$ and $b_{}^{z} \to a_{}^{z}$. This is the so-called ``optic" antiferromagnetic mode where the two antiferromagnetic sublattices oscillate out-of-phase. Eqs.~\eqref{AxEquation2}-\eqref{P2xEquation} reduce to:
  \bea
  -\frac{i \omega}{\gamma} a_{}^{x} &=& -a_{}^{y} \left( \left[ 2 H_{d}+2 H_{ex}+H_{a} \right] \sin \theta +2 H_{DM} \cos \theta \right) + M_{0} h_{}^{z} \cos \theta \nn \\
  && - a_{}^{z} \left( \left[ 2 H_{d}+2 H_{ex}+H_{a} \right] \cos \theta - 2 H_{DM} \sin \theta \right) - p_{}^{x} 2 \Gamma P_{0} M_{0}^{2} \sin (2 \theta) 
  \label{AXoptic} \\
  -\frac{i \omega}{\gamma} a_{}^{y} &=& a_{}^{x} \left( \left[ 2 H_{d} +2 H_{ex} \right] \sin \theta + H_{DM} \cos \theta \right)
  \label{AYoptic} \\
   -\frac{i \omega}{\gamma} a_{}^{z} &=& a_{}^{x} \left( H_{a} \cos \theta - H_{DM} \sin \theta \right)
   \label{AZoptic} \\
   - \frac{\omega^2}{f} p_{}^{x} &=& \left( \xi-2 \Gamma M_{0}^{2} [- \cos^2 \theta + \sin^2 \theta] - 3 \Delta P_{0}^2 \right) p_{}^{x} 
\nn \\
&& - 4 \Gamma P_{0} M_{0} \left(  - a_{}^{y} \cos \theta  + a_{}^{z} \sin \theta   \right) .
\label{Poptic}
  \eea

Eqs.~\eqref{AYoptic}-\eqref{Poptic} can be substituted into Eq.~\eqref{AXoptic} to get an equation involving only $a_{}^{x}$:
\be
a_{}^{x} = \frac{ i \omega \gamma M_{0} h_{}^{z} \cos \theta }{ \omega^2 - \omega_{o}^2 },
\ee
 where the optic antiferromagnetic frequency $\omega_{o}$ is given by:
 \bea
  \frac{ \omega_{o}^2 }{ \gamma^2 } &=& \left( \left[2 H_{d} +2 H_{ex} \right] \sin \theta + H_{DM} \cos \theta \right) \left( \left[ 2 H_{d} + 2 H_{ex} + H_{a} \right] \sin \theta +2 H_{DM} \cos \theta \right) 
\nn \\
&& +  \left( H_{a} \cos \theta - H_{DM} \sin \theta  \right) \left( \left[ 2 H_{d} + 2 H_{ex} + H_{a} \right] \cos \theta - 2 H_{DM} \sin \theta  \right) 
\nn \\
&&- \frac{8 \Gamma ^2 P_{0}^{2} M_{0}^{3} \sin (2 \theta) \left( \cos \theta \sin \theta \left[ 2 H_d +2 H_{ex} +H_a \right]-H_{DM} \right) }{\frac{\omega^2}{f} + \left( \xi-2 \Gamma M_{0}^{2} [- \cos^2 \theta + \sin^2 \theta] - 3 \Delta P_{0}^2 \right) } .
\label{OpticFreq}
 \eea
Since the Dzyaloshinskii-Moriya canting angle is small, $\omega_{o}$ is approximated very accurately by taking the limit $\sin \theta \to 0$ and $\cos \theta \to 1$. This gives:
 \be
 \frac{ \omega_{o}^2 }{ \gamma^2 } \sim 2 H_{DM}^2 + H_{a} \left( 2 H_{d} + 2 H_{ex} + H_{a} \right).
 \label{approxOpticFreq}
 \ee
 Ignoring the effective Dzyaloshinskii-Moriya field $H_{DM}$, this frequency agrees with the well-known result for thin film antiferromagnets with no canting \cite{loudonSW}. Eq.~\eqref{approxOpticFreq} also agrees with the resonant frequency calculated previously for bulk canted antiferromagnets when instead $H_{d}=0$ \cite{pincus60,MoriyaRado}.
 
 The $xz$ component of the magnetic susceptibility $\chi_{xz}^{m} = (a_{}^{x}+b_{}^{x})/h_{}^{z}$ is zero since $a_{}^{x}=-b_{}^{x}$. Similarly, $\chi_{yz}^{m}$ is zero. The nonzero susceptibility components due to $h_{}^{z}$ are $\chi_{zz}^{m} = (a^{z}+b^{z})/h^{z}$ and the electromagnetic susceptibility $\chi_{xz}^{em} = p_{}^{x} / h_{}^{z}$ which are given exactly by:
 \bea
\chi_{zz}^{m} &=& \frac{ -2 \gamma^2 M_{0} \cos \theta \left( H_{a} \cos \theta - H_{DM} \sin \theta \right) }{  \omega^2 - \omega_{o}^2 } ,
\label{WFMchizzM} \\
\chi_{xz}^{em} &=& \frac{  4 f \Gamma P_{0} \gamma^2 M_{0}^2 \cos \theta \left( \cos \theta \sin \theta \left[ 2 H_d + 2 H_{ex}-H_a \right] - H_{DM} \right) }{ \left( \omega^2 - \omega_{o}^2 \right)  \left( \omega^2- \omega_{fe}^{2} \right) } .
\label{WFMchixzEM}
 \eea
 $\chi_{xz}^{em}$ has a pole at the optic antiferromagnetic mode frequency $\omega_{o}$ and also at the ferroelectric mode frequency $\omega_{fe}$ given by
 \be
 \frac{\omega_{fe}^2}{f} = - \xi+ 2 \Gamma M_{0}^{2} [- \cos^2 \theta + \sin^2 \theta] + 3 \Delta P_{0}^2 .
 \label{FerroFreq}
 \ee
 
Driving fields $e_{}^{x}$ excite the same modes: the ferroelectric mode and the optic mode. Eqs.~\eqref{AXoptic}-\eqref{Poptic} remain the same apart from the removal of $h_{}^{z}$ and the inclusion of $e_{}^{x}$. Following the same working, it is found that $\chi_{zx}^{me} = (a_{}^{z}+b_{}^{z})/e_{}^{x} = \chi_{xz}^{em}$, which is given in Eq.~\eqref{WFMchixzEM}. The only other nonzero component appearing due to $e_{}^{x}$ is:
\bea
\chi_{xx}^{e} &=& -f \Big\{  \omega^2 -  \gamma^2 \left( \left[2 H_{d} +2 H_{ex} \right] \sin \theta + H_{DM} \cos \theta \right) \left( \left[ 2 H_{d} + 2 H_{ex} + H_{a} \right] \sin \theta +2 H_{DM} \cos \theta \right) \Big.
\nn \\
&& \Big. -  \gamma^2 \left( H_{a} \cos \theta - H_{DM} \sin \theta  \right) \left( \left[ 2 H_{d} + 2 H_{ex} + H_{a} \right] \cos \theta - 2 H_{DM} \sin \theta  \right)   \Big\} 
\nn \\
&& \Bigg/  \left( \omega^2 - \omega_{fe}^{2} \right)  \left( \omega^2 - \omega_{o}^2 \right) \\
\label{WFMchixxE}
&\sim & \frac{ -f}{ \omega^2 - \omega_{fe}^{2} } .    \nn
\eea

Next we set $h_{}^{x} \ne 0$ and $h_{}^{z}=h_{}^{y}=e_{}^{x} =0$ in Eqs.~\eqref{AxEquation2}-\eqref{P2xEquation} to solve for the susceptibility components $\chi_{ix}$ ($i=x,y,z$). The equations for $\bs{a}$ and $\bs{b}$ are symmetric under transform of $a_{}^{x} \to b_{}^{x}$, $a_{}^{y} \to b_{}^{y}$ and $a_{}^{z} \to -b_{}^{z}$, which corresponds to the antiferromagnetic sublattices oscillating in-phase and is referred to as the ``acoustic" mode. The magnetoelectric coupling term at the end of Eq.~\eqref{P2xEquation} vanishes and hence $p_{}^{x}=0$ when the system is driven by a magnetic field in the $x$ direction. 

Eqs.~\eqref{AxEquation2}-\eqref{P2xEquation} reduce to:
\bea
  -\frac{i \omega}{\gamma} a_{}^{x} &=& -a_{}^{y} \left( \left[ 2 H_{d} + H_{a} \right] \sin \theta \right)-a_{}^{z} H_{a} \cos \theta
  \label{AXacoustic} \\
  -\frac{i \omega}{\gamma} a_{}^{y} &=& a_{}^{x} \left( 2 H_{d} \sin \theta + H_{DM} \cos \theta \right) + M_{0} \sin \theta h_{}^{x}
  \label{AYacoustic} \\
   -\frac{i \omega}{\gamma} a_{}^{z} &=& a_{}^{x} \left( \left[2 H_{ex} + H_{a} \right] \cos \theta - H_{DM} \sin \theta \right) - M_{0} \cos \theta h_{}^{x} .
   \label{AZacoustic}    
\eea

The two nonzero components of the susceptibility $\chi_{xx}^{m}$ and $\chi_{yx}^{m}$ are given by:
\bea
\chi_{xx}^{m} &=& \frac{ 2 \gamma^2 M_{0} \left( \sin^2 \theta \left( 2 H_{d} + H_{a} \right) - \cos^2 \theta H_{a} \right)  }{ \omega^2 - \omega_{a}^2 }
\label{WFMchixxM} \\
\chi_{yx}^{m} &=& \frac{ i 2 \gamma M_{0} \left( \omega^2 \sin \theta - \gamma^2 H_{a} \cos \theta \left(  \left[ 2 H_{d} + 2 H_{ex} + H_{a} \right] \cos \theta \sin \theta  - H_{DM}  [1 -2\cos^2 \theta ]  \right) \right) }{ \omega \left( \omega^2 - \omega_{a}^{2} \right) } ,
\label{WFMchiyxM}
\eea
where the acoustic antiferromagnetic resonant frequency is
\be
\frac{ \omega_{a}^{2} }{ \gamma^2 } = H_{a} \left( 2 H_{ex} + H_{a} \right) \cos^2 \theta + H_{d} H_{DM} \sin (2 \theta) + 2 H_{d} \left(2 H_{d} + H_{a} \right) \sin^2 \theta.
\label{AcousticFreq}
\ee
Once again, if we make the approximation $\sin \theta \to 0$, then this expression reduces to the known acoustic frequency (whether there is canting or not) given by \cite{loudonSW, pincus60,MoriyaRado}
\be
\frac{ \omega_{a}^{2} }{ \gamma^2 } \sim H_{a} \left( 2 H_{ex} + H_{a} \right) .
\label{approxAcousticFreq}
\ee

Making $h_{}^{y} \ne 0$ and $h_{}^{x} = h_{}^{z} = e_{}^{x} =0$, we find that driving fields in the $y$~direction also excite the acoustic antiferromagnet mode and do not excite a dielectric mode. By symmetry, the susceptibility component $\chi_{xy}^{m} = - \chi_{yx}^{m}$ and so has already been found. The remaining susceptibility component $\chi_{yy}^{m}$ is found to be:
\be
\chi_{yy}^{m} =\frac{2 \gamma^2 M_{0} \sin \theta \left( 2 H_{d} \sin \theta + H_{DM} \cos \theta  \right) } {  \omega^2 - \omega_{a}^{2} } .
\label{WFMchiyyM}
\ee
This component vanishes as $\theta \to 0$ since then the linearization of the antiferromagnetic sublattice magnetizations requires that there is no dynamic magnetization in the $y$~direction.

Finally, there are two non-zero components of the electric susceptibility given by examining Eqs.~\eqref{Pyequation}-\eqref{Pzequation}:
\bea
\chi_{yy}^{e} &=& \frac{f}{ -\omega^2 + f \xi_{\perp} }
\label{WFMchiyyE}
\\
\chi_{zz}^{e} &=& \frac{f}{-\omega^{2} + f (\xi_{\perp}+4 \pi) } .
\label{WFMchizzE}
\eea

The total susceptibility tensor for the ferroelectric weak ferromagnet geometry takes the form:
\be
\hat{\chi} = \left(  \begin{array}{cccccc}    \chi_{xx}^{m} & \chi_{xy}^{m} & 0 & 0 & 0 & 0 \\    \chi_{yx}^{m} & \chi_{yy}^{m} & 0 & 0 & 0 & 0 \\   0 & 0 & \chi_{zz}^{m} & \chi_{zx}^{me} & 0 & 0      \\   0  & 0 &  \chi_{xz}^{em} & \chi_{xx}^{e} & 0 & 0 \\   0 & 0 & 0 & 0 & \chi_{yy}^{e} & 0   \\  0 & 0 & 0 & 0& 0 & \chi_{zz}^{e}    \end{array} \right) ,
\label{susc_form_canting}
\ee
with the eight independent components being given in Eqs.~\eqref{WFMchizzM}, \eqref{WFMchixzEM}, \eqref{WFMchixxE}, \eqref{WFMchixxM}, \eqref{WFMchiyxM}, \eqref{WFMchiyyM}, \eqref{WFMchiyyE} and \eqref{WFMchizzE}.

 \section{Weak ferromagnet/ferromagnet heterostructure}
 \label{heterostructure}
 
 \subsection{Energy density and equations of motion}
 
We now consider a heterostructure comprised of alternating thin films of a ferroelectric weak ferromagnet, as illustrated in Fig.~\ref{WFMgeom}, with thickness $d_{w}$ and a ferromagnet with thickness $d_{f}$ in the $z$-direction. We shall assume that the ferromagnet has a uniaxial anisotropy in the $y$~direction with strength $K_{f}$ and dielectric stiffness components given by $\alpha_{x}$, $\alpha_{y}$ and $\alpha_{z}$. Then the energy density of the ferromagnetic film is given by:
\be
\mathscr{E}_{FM} = - K_{f} (M_{f}^{y})^2 - \bs{H}_{f} \cdot \bs{M}_{f} + \frac{1}{2} \sum_{i=x,y,z} \alpha_{i} (p_{f}^{i})^2 - \bs{E}_{f} \cdot \bs{p}_{f},
\label{FMhetero}
\ee
where $\bs{M}_{f} = (m_{}^{x}, M_{f} + m_{}^{y}, m_{}^{z} )$ is the linearized magnetization and $\bs{p}_{f}$ is the dynamic dielectric polarization and so is denoted by a lower case letter. $\bs{H}_{f}$ and $\bs{E}_{f}$ are the dipolar magnetic and electric fields respectively. They have been written in upper case to emphasize that these may have a static as well as a dynamic part.

We rewrite the energy density for the ferroelectric weak ferromagnet shown in Eqs.~\eqref{newAFMenergy} and \eqref{E_FElayer} so that the thin film demagnetizing and depolarizing terms are discarded and the dipolar fields are written in a corresponding way as to in the ferromagnet:
\bea
\mathscr{E}_{AFM} &=& (J+ \Gamma (P^{x})^{2} ) \bs{M}_{a} \cdot \bs{M}_{b} - K \left[ \left(M_{a}^{y} \right)^2 + \left( M_{b}^{y} \right)^2  \right] 
\nn \\
&&+ D (M_{a}^{y} M_{b}^{z} - M_{a}^{z} M_{b}^{y} ) - \bs{H}_{w} \cdot \bs{M} 
\label{AFMhetero} \\
\mathscr{E}_{FE} &=& - \frac{1}{2} \xi \left( P^{x} \right)^2 + \frac{1}{4} \Delta \left( P^{x} \right)^4  - \bs{E}_{w} \cdot \bs{P}  \nn \\
&&+ \frac{1}{2} \xi_{\perp} \left[ \left( P^{y} \right)^2 + \left( P^{z} \right)^2 \right]  .
\label{FEhetero}
\eea

To calculate the analytic susceptibility and the resonant $k=0$ frequencies of the heterostructure, we use an effective medium method which requires that Maxwell's boundary conditions for dipole fields are satisfied at the interfaces between the materials \cite{agranovich,raj87, almeida88}. This method can also be used to numerically calculate the frequencies of long wavelength spin waves with finite wavevector ($k \ne 0$) in an approach known as entire-cell effective medium method \cite{bob96, me06} and gives results in good agreement with more computational-demanding methods for including dipolar interactions.

Maxwell's boundary conditions relate the dipolar fields in the ferromagnet ($\bs{H}_{f}$, $\bs{B}_{f}=\bs{H}_{f}+ 4 \pi \bs{M}_{f}$, $\bs{E}_{f}$ and $\bs{D}_{f}= \bs{E}_{f}+4 \pi \bs{P}_{f}$) and the weak ferromagnet ($\bs{H}_{w}$, $\bs{B}_{w}=\bs{H}_{w}+ 4 \pi (\bs{M}_{a} + \bs{M}_{b})$, $\bs{E}_{w}$ and $\bs{D}_{w}= \bs{E}_{w}+4 \pi \bs{P}_{w}$) according to:
\bea
H_{f}^{x} &=& H_{w}^{x} = h_{}^{x}
\label{HxBC} \\
H_{f}^{y} &=& H_{w}^{y}  = h_{}^{y}\\
H_{f}^{z} + 4 \pi M_{f}^{z} &=& H_{w}^{z} + 4 \pi (M_{a}^{z} +M_{b}^{z} )= C 
\label{BzBC} \\
E_{f}^{x} &=& E_{w}^{x} = e_{}^{x} \\
E_{f}^{y} &=& E_{w}^{y} = e_{}^{y} \\
E_{f}^{z} + 4 \pi P_{f}^{z} &=& E_{w}^{z} + 4 \pi P_{w}^{z} = D , \label{DzBC}
\eea
for the geometry shown in Fig.~\ref{WFMgeom}. The constants $C$ and $D$ are defined for ease of notation in what follows. In particular, the out-of-plane ($z$) boundary conditions couple the dipolar fields to the magnetization and electric polarization in both materials. For this reason we must calculate the $iz$ components of the susceptibility tensor first in order to properly take into account dipolar effects.

All of the dipolar fields are in fact dynamic, apart from $H_{w}^{z}$ since from the linearization Eqs.~\eqref{MaLinearise} and \eqref{MbLinearise} together with the boundary condition Eq.~\eqref{BzBC}
\be
H_{w}^{z} = C - 8 \pi M_{0} \sin \theta - 4 \pi ( a_{}^{z} + b_{}^{z} ) .
\ee
It is the dynamic part of $H_{w}^{z}$, namely $h_{w}^{z}=C-4 \pi ( a_{}^{z} + b_{}^{z} )$, which drives the magnetization and which we need to find in order to calculate the dynamic effective medium susceptibility.

Substituting the energy densities Eqs.~\eqref{FMhetero}-\eqref{FEhetero} and boundary conditions Eqs.~\eqref{HxBC}-\eqref{DzBC} into the equations of motion Eqs.~\eqref{LLequation} and \eqref{Peqnmotion} we obtain the following equations. For the ferromagnet we have:
\bea
-\frac{i \omega}{\gamma} m_{}^{x} &=& -m_{}^{z} (H_{af} + H_{df} )  + M_{f} C
\label{M1xEquation}  \\
-\frac{i \omega}{\gamma} m_{}^{z} &=&  m_{}^{x}  H_{af} +M_{f} h_{}^{x}  ,
\label{M1zEquation}
\eea
where the effective anisotropy and static dipolar fields in the ferromagnet are given by $H_{af}=2 K_{f} M_{f}$ and $H_{df}=4 \pi M_{f}$ respectively. We assume that the gyromagnetic ratio $\gamma$ is the same for both the ferromagnet and the weak ferromagnet.

The equations of motion for $\bs{a}$ and $\bs{b}$ are the same as shown for the weak ferromagnet thin film in Eqs.~\eqref{AxEquation2}-\eqref{BzEquation2} apart from the replacement of $h_{}^{z} \to C$. This is deceptive as it appears that the boundary conditions have simply created a thin film demagnetizing effect. However, since the driving field is $h_{w}^{z}=C-4 \pi ( a_{}^{z} + b_{}^{z} )$ rather than $C$, this is not the case as will be shown later.

The linearized electric equations of motion are:
\bea
- \frac{\omega^2}{ f_f} p_{f}^{x} &=& - \alpha_{x} p_{f}^{x} + e_{}^{x} 
\label{P1xEquation2} \\
- \frac{\omega^2}{f_f} p_{f}^{y} &=& - \alpha_{y} p_{f}^{y} + e_{}^{y} 
\label{P1yEquation2} \\
- \frac{\omega^2}{f_f} p_{f}^{z} &=& - (\alpha_{z}+4 \pi) p_{f}^{z} + D
\label{P1zEquation2} 
\eea
\bea
- \frac{\omega^2}{f_w} p_{w}^{x} &=& \left( \xi-2 \Gamma M_{2}^{2} [- \cos^2 \theta + \sin^2 \theta]-2 K_{ME} M_{1}^{2} - 3 \Delta P_{0}^2 \right) p_{w}^{x} + e_{}^{x} 
\nn \\
&& - 2 \Gamma P_{0} M_{2} \left(  - a_{}^{y} \cos \theta  +  b_{}^{y} \cos \theta + a_{}^{z} \sin \theta + b_{}^{z} \sin \theta   \right) ,
\label{P2xEquation2} \\
- \frac{\omega^2}{f_w} p_{w}^{y} &=& - \xi_{\perp} p_{w}^{y} + e_{}^{y} 
\label{P2yEquation2} \\
- \frac{\omega^2}{f_w} p_{w}^{z} &=& - (\xi_{\perp} +4 \pi) p_{w}^{z} + D ,
\label{P2zEquation2}
\eea
where $f_f$ and $f_w$ are the effective inverse mass terms of the respective dielectric materials.

\subsection{Effective medium susceptiility}

As shown in Sec.~\ref{SinglePhase}, the magnetizations and electric polarizations can be found as a function of dipolar field and then the susceptibility can be derived. As already mentioned, the $\chi_{iz}$ ($i=x,y,z$) components of the susceptibility must be found first for the heterostructure which involves setting $C \ne 0$ and $D\ne0$ and ignoring all other dipolar field components. In the effective medium approximation the fields in the two materials are averaged (for example, $\langle m_{}^{x} \rangle = d_{f} m_{}^{x}  + d_{w} ( a_{}^{x}+b_{}^{x})$) giving susceptibility components:
\bea
\chi_{iz}^{m} &=& \frac{ d_{f} m_{}^{i}  + d_{w} ( a_{}^{i}+b_{}^{i} ) }{ d_{f} (C- 4 \pi m_{}^{z}) + d_{w} (C- 4 \pi a_{}^{z} - 4 \pi b_{}^{z} ) }  \equiv \frac{ \langle m_{}^{i} \rangle }{ \langle h_{}^{z} \rangle }, 
\label{Chi_izM} \\
\chi_{iz}^{e} &=& \frac{ d_{f}^{} p_{f}^{i}  + d_{w}^{} p_{w}^{i} }{ d_{f} (D- 4 \pi p_{f}^{z}) + d_{w} (D- 4 \pi p_{w}^{z}) } \equiv \frac{ \langle p_{}^{i} \rangle }{ \langle e_{}^{z} \rangle } ,
\label{Chi_izE} \\
\chi_{iz}^{em} &=& \frac{ d_{f}^{} p_{f}^{i}  + d_{w}^{} p_{w}^{i} }{ d_{f} (C- 4 \pi m_{}^{z}) + d_{w} (C- 4 \pi a_{}^{z} - 4 \pi b_{}^{z} ) } \equiv \frac{ \langle p_{}^{i} \rangle }{ \langle h_{}^{z} \rangle } ,
\label{Chi_izEM} \\
\chi_{iz}^{me} &=& \frac{ d_{f}^{} m_{}^{i}  + d_{w}^{} ( a_{}^{i}+b_{}^{i} ) }{ d_{f} (D- 4 \pi p_{f}^{z}) + d_{w} (D- 4 \pi p_{w}^{z}) } \equiv \frac{ \langle m_{}^{i} \rangle }{ \langle e_{}^{z} \rangle }  .
\label{Chi_izME}
\eea
Each term is weighted by the corresponding film thickness $d_{w}$ or $d_{f}$.

The other susceptibility components can then be found. For example, by setting $h_{}^{x} \ne0$ the $\chi_{ix}^{m}$ and $\chi_{ix}^{em}$ components can be found according to:
\bea
\chi_{ix}^{m} &=&  \frac{  d_{f}^{} m_{}^{i} + d_{w} (a_{}^{i}+b_{}^{i})  - \chi_{iz}^{m} \langle h_{}^{z} \rangle   - \chi_{iz}^{me} \langle e_{}^{z} \rangle   }{  (d_{f}^{}+d_{w}^{}) h_{}^{x}  }
\nn \\
& \equiv& \frac{ \langle m_{}^{i} \rangle - \chi_{iz}^{m} \langle h_{}^{z} \rangle   - \chi_{iz}^{me} \langle e_{}^{z} \rangle   }{ \langle h_{}^{x} \rangle },
\label{Chi_ixM} \\
\chi_{ix}^{em} &=&\frac{ \langle p_{}^{i} \rangle - \chi_{iz}^{em} \langle h_{}^{z} \rangle   - \chi_{iz}^{e} \langle e_{}^{z} \rangle   }{ \langle h_{}^{x} \rangle },
\label{Chi_ixEM}
\eea
where the susceptibility components on the right hand side of Eqs.~\eqref{Chi_ixM} and \eqref{Chi_ixEM} were found in the previous step using Eqs.~\eqref{Chi_izM}-\eqref{Chi_izME}.

It should be noted that the method described is identical to existing effective medium methods \cite{agranovich,raj87, almeida88,bob96,me06}, apart from the fact that we treat \emph{both} dielectric and magnetic systems for the first time. This only works since the equation of motion for $p_{w}^{z}$ and $p_{f}^{z}$ are not coupled to the equation of motion for $m_{}^{z}$, $a_{}^{z}$ and $b_{}^{z}$. If the out-of-plane dielectric and magnetic oscillations were coupled, then the effective dipolar fields $\langle h_{}^{z} \rangle$ and $\langle e_{}^{z} \rangle$ would be coupled and much more complicated expressions for the susceptibility components, as compared with Eqs.~\eqref{Chi_izM}-\eqref{Chi_izME}, would need to be found. This represents a new extension to the effective medium method which will be discussed in a later paper.

Without providing working, the result for the non-zero components of the frequency-dependent susceptibility is
\bea
\chi_{zz}^{m} &=& \frac{  -d_{f} \gamma^2 M_{f}^{} H_{af} ( \omega^2- \omega_{o}^2 ) - d_{w}^{} \gamma^2 M_{0} \cos \theta \left( H_{a} \cos \theta - H_{DM} \sin \theta \right) \left( \omega^2 - \omega_{f}^2 \right) }{  d_{f}^{} \left( \omega^2 - \omega_{o}^2 \right) \left( \omega^2 - \omega_{f\ast}^{2}\right)  +d_{w} \left( \omega^2 - \omega_{o\ast}^2 \right) \left( \omega^2 - \omega_{f}^{2}\right) }
\label{ChiZZm-hetero} 
\\
\chi_{xz}^{m} &=& \frac{  -i \omega \gamma M_{f} d_{f} ( \omega^2- \omega_{o}^2 )  }{  d_{f}^{} \left( \omega^2 - \omega_{o}^2 \right) \left( \omega^2 - \omega_{f\ast}^{2}\right)  +d_{w} \left( \omega^2 - \omega_{o\ast}^2 \right) \left( \omega^2 - \omega_{f}^{2}\right)  }
\label{ChiXZm-hetero}
\\
\chi_{yx}^{m} &=& \frac{ i 2 \gamma M_{0} d_{w} \left( \omega^2 \sin \theta - \gamma^2 H_{a} \cos \theta \left(  \left[ 2 H_{d} + 2 H_{ex} + H_{a} \right] \cos \theta \sin \theta  - H_{DM}  [1 -2\cos^2 \theta ]  \right) \right) }{ \omega (d_{f}^{}+d_{w}^{}) \left( \omega^2 - \omega_{a}^{2} \right) }
\label{ChiYXm-hetero}
\\
\chi_{xx}^{m} &=& \frac{  d_{f}^{} M_{f}^{} \gamma^2  \left\{  d_{f}^{} H_{af}^{} \left( \omega^2 - \omega_{o}^2 \right)  +  d_{w} \left( H_{af}^{} + 4 \pi M_{f}^{} \right)  \left( \omega^2 - \omega_{o\ast}^2 \right)  \right\}   }{  \left(d_{f}^{}+d_{w}^{} \right)   \left\{ d_{f}^{} \left( \omega^2 - \omega_{o}^2 \right) \left( \omega^2 - \omega_{f\ast}^{2}\right)  +d_{w} \left( \omega^2 - \omega_{o\ast}^2 \right) \left( \omega^2 - \omega_{f}^{2}\right) \right\}  }
\nn \\
&& +\frac{  2 d_{w}^{} \gamma^2 \left( M_{0}^{} \sin^2 \theta (2 H_{d} + H_{a} ) - M_{0}^{} H_{a}^{} \cos^2 \theta \right)  }{  \left( d_{f}+d_{w} \right)  \left(\omega^2 -\omega_{a}^{2} \right) }
\label{ChiXXm-hetero}
\\
\chi_{yy}^{m} &=& \frac{2 d_{w}^{} \gamma^2 M_{0} \sin \theta \left( 2 H_{d} \sin \theta + H_{DM} \cos \theta  \right) } {  \left( d_{f} + d_{w} \right)  \left( \omega^2 - \omega_{a}^{2} \right) }
\label{ChiYYm-hetero}
\\
\chi_{xz}^{em} &=& \frac{ -d_{w} 4 \Gamma \gamma^2 f_{w}^{} P_{0} M_{0}^{2} \cos \theta \left( \cos \theta \sin \theta [ 2 H_{d}+2 H_{ex}-H_{a}] +H_{DM}  \right) \left(  \omega^2- \omega_{f}^{2}  \right)   }{ \left( \omega^2 - \omega_{fe}^{2} \right)  \left\{ d_{f}^{} \left( \omega^2 - \omega_{o}^2 \right) \left( \omega^2 - \omega_{f\ast}^{2}\right)  +d_{w} \left( \omega^2 - \omega_{o\ast}^2 \right) \left( \omega^2 - \omega_{f}^{2}\right) \right\} }
\label{ChiXZem-hetero}
\\
\chi_{xx}^{em} &=& \frac{  i \omega d_{f} d_{w} 4 \pi \gamma^3 4 \Gamma f_{w}^{}P_{0} M_{0}^2  M_{f}  \cos \theta \left( \cos \theta \sin \theta [ 2 H_{d}+2 H_{ex}-H_{a}] +H_{DM}  \right) }{  (d_{f}^{}+d_{w}^{}) \left( \omega^2 - \omega_{fe}^{2} \right)  \left\{ d_{f}^{} \left( \omega^2 - \omega_{o}^2 \right) \left( \omega^2 - \omega_{f\ast}^{2}\right)  +d_{w} \left( \omega^2 - \omega_{o\ast}^2 \right) \left( \omega^2 - \omega_{f}^{2}\right) \right\}  }
\label{ChiXXem-hetero}
\\
\chi_{zz}^{e} &=& \frac{  d_{f} \left( - {\omega^2}/{f_w} + \xi_{\perp} + 4 \pi \right)+ d_{w} \left( - {\omega^2}/{f_{f}} + \alpha_{z} + 4 \pi \right)   }{  d_{f} \left( {\omega^2}/{f_f} - \alpha_{z} \right)  \left(  {\omega^2}/{f_w} - \xi_{\perp} - 4 \pi \right)+ d_{w}  \left( {\omega^2}/{f_w} - \xi_{\perp} \right) \left(  {\omega^2}/{f_{f}} - \alpha_{z} - 4 \pi \right)   }
\label{ChiZZe-hetero}
\\
\chi_{xx}^{e} &=&  \frac{-1}{(d_{f}^{}+d_{w}^{} )} \left(   \frac{d_{f}^{} f_{f}^{} }{{\omega^2} - f_{f} \alpha_{x}}  +  \frac{d_{w}^{} f_{w}^{}}{{\omega^2}- \omega_{fe}^2}  \right)
\label{ChiXXe-hetero}
\\
\chi_{yy}^{e} &=& \frac{-1}{(d_{f}^{} +d_{w}^{}) } \left(   \frac{d_{f}^{} f_{f}^{}}{{\omega^2} -f_{f}  \alpha_{y}}  +  \frac{d_{w}^{} f_{w}^{}}{{\omega^2}-  f_{w}^{}\xi_{\perp}  }  \right),
\label{ChiYYe-hetero}
\eea
where the optic antiferromagnetic frequency in a weak ferromagnetic thin film $\omega_{o}^{}$ is given in Eq.~\eqref{OpticFreq}, the acoustic antiferromagnetic frequency $\omega_{a}^{}$ is given in Eq.~\eqref{AcousticFreq} and the ferroelectric mode frequency $\omega_{fe}^{}$ is given in Eq.~\eqref{FerroFreq}. In addition, we have new frequencies for the ferromagnet in thin film $\omega_{f}^{}$ and in bulk $\omega_{f\ast}^{}$:
\bea
\frac{\omega_{f}^{2} }{\gamma^2}  &=& H_{af}^{} \left( H_{af}^{} + 4 \pi M_{f}^{} \right),
\\
\frac{\omega_{f\ast}^{2} }{\gamma^2}  &=& H_{af}^{2} .
\eea
An additional frequency associated with the optic antiferromagnetic mode in bulk is given by
\be
\frac{ \omega_{o\ast}^{2} }{\gamma^2} = \frac{ \omega_{o}^{2} }{\gamma^2} - 2 H_{d}^{} \cos \theta \left( H_{a} \cos \theta - H_{DM} \sin \theta \right) .
\ee

Compared with the weak ferromagnet in isolation (Sec.~\ref{SinglePhase}) two extra components are non-zero in the susceptibility tensor, namely $\chi_{xz}^{m}$ and $\chi_{xx}^{em}$. $\chi_{xz}^{m}$ appears since it is non-zero in the ferromagnet. $\chi_{xx}^{em}$ arises purely due to the coupling of $m_{}^{x}$ in the ferromagnet to the out-of-plane dipolar field and is given by:
\be
\chi_{xx}^{em} = \frac{  d_{f} 4 \pi m_{}^{x} \chi_{xz}^{em}  }{  (d_{f}^{}+d_{w}^{} ) h_{}^{x}  } .
\ee

\subsection{Limiting cases}

We consider some limiting cases to test the effective medium method. We use $\chi_{zz}^{m}$ (see Eq.~\eqref{ChiZZm-hetero}) to demonstrate the results.

First we consider replacing the ferromagnet with a nonmagnetic material ($M_{f}^{} \to 0$). The component becomes
\be
\chi_{zz}^{m} = \frac{ - d_{w} \gamma^2 M_{0} \cos \theta \left( H_{a}^{} \cos \theta - H_{DM}^{} \sin \theta \right) }{ d_{f}^{} \left( \omega^{2} - \omega_{o}^2 \right)+  d_{w}^{} \left( \omega^{2} - \omega_{o\ast}^{2} \right) } .
\ee
Then taking the limit that the weak ferromagnetic films are much thinner than the nonmagnetic spacers ($d_{f}^{} >> d_{w}^{}$), the isolated thin film result from Sec.~\ref{SinglePhase} is recovered (Eq.~\eqref{WFMchizzM}), namely
\be
\chi_{zz}^{m,\textrm{film}} = \frac{ -2 \gamma^2 M_{0} \cos \theta \left( H_{a} \cos \theta - H_{DM} \sin \theta \right) }{  \omega^2 - \omega_{o}^2 } .
\ee

Next we consider removing the ferromagnetic layers ($d_{f} \to 0$). This gives,
\be
\chi_{zz}^{m,\textrm{bulk}} = \frac{ -2 \gamma^2 M_{0} \cos \theta \left( H_{a} \cos \theta - H_{DM} \sin \theta \right) }{  \omega^2 - \omega_{o\ast}^2 } 
\ee
which is the same as the result found for the isolated thin film weak ferromagnet (Eq.~\eqref{WFMchizzM}), apart from the pole being at the bulk frequency $\omega_{o\ast}^{}$ rather than at the thin film frequency $\omega_{o}^{}$. Hence the bulk limit is correctly recovered. Similarly, the bulk ferromagnetic susceptibility is recovered in the limit of the weak ferromagnet vanishing:
\be
\chi_{zz}^{m,\textrm{bulk}} = \frac{  - \gamma^2 H_{af} M_{f}  }{  \omega^2 - \omega_{f\ast}^{2}  } .
\ee

The effective medium method recovers the correct limits for both bulk and thin film geometries and therefore seems reliable.

\subsection{Dynamic magnetoelectric coupling}

What is most significant when examining Eqs.~\eqref{ChiZZm-hetero}-\eqref{ChiYYe-hetero} is that the effective medium susceptibility is not, in general, given by an average of the susceptibility in each film. This means that instead of finding poles in $\chi_{zz}^{m}$ at the ferromagnetic bulk frequency $\omega_{f\ast}^{}$ and at the optic antiferromagnetic bulk frequency $\omega_{o\ast}^{}$, we find two resonant frequencies given by the solution to:
\be
0=d_{f}^{} \left( \omega^2 - \omega_{o}^2 \right) \left( \omega^2 - \omega_{f\ast}^{2}\right)  +d_{w} \left( \omega^2 - \omega_{o\ast}^2 \right) \left( \omega^2 - \omega_{f}^{2}\right).
\label{eqforFreq}
\ee
These two frequencies correspond to modes that are common to both materials and are a signature of the fact that the out-of-plane dipolar magnetic fields serve to hybridize the ferromagnetic and optic antiferromagnetic resonances. In Sec.~\ref{SinglePhase} we showed that for a weak ferromagnet with magnetostrictive magnetoelectric coupling, the ferroelectric and optic antiferromagnetic modes are coupled. This in turn means that the ferromagnetic resonance is coupled to the ferroelectric mode. Examining the expression for $\chi_{xz}^{em}$ in Eq.~\eqref{ChiXZem-hetero}, it is indeed seen that there is a magnetoelectric resonance involving the ferromagnet.

To demonstrate that one of the magnetoelectric resonant frequency may be in the GHz regime through this indirect coupling of ferroelectric-optic-ferromagnetic modes, approximate frequencies for a NiFe/BiFeO$_3$ (ferromagnet/weak ferromagnet) heterostructure are calculated. Equal volumes of both materials are assumed ($d_{f}^{}=d_{w}^{}$). The relevant frequencies of the isolated films and bulk samples are given in Table~\ref{freqs}. Substituting these into Eq.~\eqref{eqforFreq}, we find two of the three magnetoelectric resonant frequencies in the heterostructure at 4.07~GHz and 548.0~GHz. The former value shows how such a heterostructure may be designed to give dynamic magnetoelectric coupling in the microwave regime. With a change in the ferromagnet used, application of an applied magnetic field, or a change in the relative thicknesses of the two materials, this frequency can be tuned.
\begin{table}[h]
\caption{\label{freqs}The resonant frequencies of NiFe (thin film and bulk ferromagnetic modes) and BiFeO$_3$ (thin film and bulk optic modes). These are estimated by assuming $\gamma=2 \pi \times 2.8 \times 10^6$~Hz/Oe, that for NiFe $H_{af}^{}=10$~Oe and $M_{f}^{}=867$~Oe, and that for BiFeO$_3$, $H_{a}^{}=880$~Oe \cite{Bai05}, $H_{ex}^{}=2.7 \times 10^5$~Oe \cite{Kadomtseva}, $M_{0}=750$~Oe \cite{wang03} and $H_{DM}^{}=1400$~Oe. The value for $H_{DM}^{}$ is estimated using the canting angle $\theta =0.14^{\circ}$ \cite{bea07} together with Eq.~\eqref{ThetaEquilib}.}
\begin{ruledtabular}
\begin{tabular}{cccc}
$\omega_{f}$	& $\omega_{f\ast}$	& $\omega_{o} $ & $\omega_{o\ast}$  \\
\hline
5.81		& 0.176				& 552.6 		& 543.3 		   \\
\end{tabular}
\end{ruledtabular}
\end{table}

We should stress that the only mechanism in this model for a dynamic magnetoelectric coupling between the weak ferromagnet and the ferromagnet is through dipolar fields. In a real system exchange coupling at the film interfaces may also lead to a dynamic magnetoelectric coupling by coupling the ferromagnetic and optic antiferromagnetic modes. Exchange coupling leads to an asymmetry between the two antiferromagnetic sublattices and so the effective medium susceptibility must be found numerically rather than analytically. Also, for relatively thick films, the exchange coupling only represents a small contribution to the energy density and so will not change the resonant frequencies significantly from those calculated here.

\section{Conclusion}
\label{conclusion}

In this work we have shown that a magnetostrictive magnetoelectric coupling together with a canting of antiferromagnetic sublattices (known as weak ferromagnetism) in a material leads to a dynamic coupling between ferroelectric and optic antiferromagnetic excitations. Such a model is applicable to known multiferroic materials, such as BiFeO$_3$. Hybrid magnetoelectric excitation (or electromagnons) are interesting for probing the origin and strength of magnetoelectric coupling but also may have application to high frequency signal processing. Most antiferromagnetic and ferroelectric resonant frequencies are in the infrared regime.

In the second part of the work, we used an existing effective medium method to calculate the high frequency susceptibility of a ferroelectric weak ferromagnet/ferromagnet heterostructure. This is the first time dielectric and magnetic susceptibilities have been found simultaneously using this method. The main result is that the magnetic dipolar coupling between the films mediates a dynamic coupling between the ferromagnetic and ferroelectric modes. This means that there is an electromagnon in the low GHz or microwave regime. Heterostructures may be designed to produce electromagnons in a desired frequency range.

The strength of the dynamic magnetoelectric coupling is in general weak via this mechanism. In fact, for applications it appears that magnetostrictive/piezoelectric composites with an interface strain-mediated coupling are much more promising since they have magnetoelectric coupling strengths up to 100 times larger than in single-phase materials (see, for example, the review article by Nan \emph{et al.} \cite{NanReview}). Such heterostructures may be similarly treated using an effective medium method, with an appropriately chosen magnetoelectric coupling between films. A microscopic entire-cell effective medium method may prove more useful for calculating the susceptibility since the magnetoelectric coupling is an interface effect.

\begin{acknowledgments}
The authors acknowledge support from the Australian Research Council. K.L.L. also acknowledges support from the Hackett Student Fund at UWA, UWA Completion Scholarship and Seagate Technologies.
\end{acknowledgments}



 \end{document}